# Magnetization dynamics and spin pumping induced by standing elastic waves

A. V. Azovtsev and N. A. Pertsev

*Ioffe Institute, 194021 St. Petersburg, Russia*

The magnetization dynamics induced by standing elastic waves excited in a thin ferromagnetic film is described with the aid of micromagnetic simulations taking into account the magnetoelastic coupling between spins and lattice strains. Our calculations are based on the numerical solution of the Landau-Lifshitz-Gilbert equation comprising the damping term and the effective magnetic field with all relevant contributions resulting from the exchange interaction, magnetocrystalline anisotropy, magnetoelastic coupling, and dipole-dipole magnetostatic interactions. The simulations have been performed for the 2 nm thick $Fe_{81}Ga_{19}$ film dynamically strained by longitudinal and transverse standing waves with various frequencies, which span a wide range around the resonance frequency $\nu_{res}$ of coherent magnetization precession in unstrained $Fe_{81}Ga_{19}$ film. It is found that standing elastic waves give rise to complex local magnetization dynamics and spatially inhomogeneous dynamic magnetic patterns. The spatio-temporal distributions of the magnetization oscillations in standing elastic waves have the form of *standing spin waves* with the same wavelength. Remarkably, the amplitude of magnetization precession does not go to zero at the nodes of these spin waves, which cannot be precisely described by simple analytical formulae. In the steady-state regime, the magnetization oscillates with the frequency of elastic wave, except for the case of longitudinal waves with frequencies well below $\nu_{res}$, where the magnetization precesses with a *variable frequency* strongly exceeding the wave frequency. Importantly, the precession amplitude at the antinodes of standing spin waves strongly increases when the frequency of elastic wave becomes close to $\nu_{res}$. The results obtained for the magnetization dynamics driven by elastic waves are used to calculate the spin current pumped from the dynamically strained ferromagnet into adjacent paramagnetic metal. The numerical calculations demonstrate that the transverse charge current in the paramagnetic layer, which is created by the spin current via the inverse spin Hall effect, is high enough to be measured experimentally.

## I. INTRODUCTION

The magnetization dynamics in ferromagnets is usually excited and controlled by magnetic fields or spin-polarized electric currents [1-4]. However, these methods are generally associated with high energy losses, which make them unsuitable for applications in advanced spintronic devices aimed at low power consumption. Therefore, intensive research efforts are currently focused on the development of alternative excitation techniques, such as the exploitation of elastic waves and strain pulses to induce the magnetization precession and switching in ferromagnets [5-8]. This "acoustic spintronics" [6], a promising emerging direction in the modern physics of ferromagnets, is based on the magnetoelastic coupling between spins and lattice strains [9], which leads to a variety of interesting physical phenomena. In particular, recent experimental studies revealed dynamic modulations of the magnetization direction by picosecond acoustic pulses [5], spin pumping via the injection of sound waves into a ferromagnetic film [6], excitation of a ferromagnetic resonance by surface acoustic waves in a ferromagnetic-ferroelectric hybrid [7], and generation of spin currents at the acoustic resonance [8].

Although the magnetic dynamics driven by elastic waves and strain pulses is inevitably spatially inhomogeneous, this important feature was either ignored in the theoretical studies [7, 10] or described for special situations in the approximation of small deviations from the equilibrium magnetization direction [11-13]. In this work, we employed micromagnetic simulations taking into account both magnetoelastic coupling and exchange interaction to describe the inhomogeneous magnetization dynamics excited by standing elastic waves generated in a thin ferromagnetic film. Computations were carried out for longitudinal and transverse waves in a galfenol film because Fe-Ga alloys have very high magnetoelastic coefficients [14]. The results were used to calculate the spin current pumped from the dynamically strained ferromagnet into adjacent paramagnetic metal. It should be emphasized that standing elastic waves can be generated in plate-like magnetic crystals by femtosecond laser pulses and were found to induce unusual magnetization dynamics [15].

## II. MICROMAGNETIC SIMULATIONS OF STRAIN-DRIVEN MAGNETIZATION DYNAMICS

Our approach is based on the numerical integration of the Landau-Lifshitz-Gilbert (LLG) torque equation describing the temporal evolution of the local magnetization $\mathbf{M}(t)$. In the considered case of highly magnetostrictive materials, the magnetization dynamics driven by elastic waves can be described by the conventional LLG equation [16], which may be written as $d\mathbf{M}/dt = -\gamma \mathbf{M} \times \mathbf{H}_{eff} + (\alpha/M_s)\mathbf{M} \times d\mathbf{M}/dt$, where $\gamma$ is the gyromagnetic ratio, $\alpha$ is the dimensionless Gilbert damping parameter, $M_s = |\mathbf{M}|$ is the saturation magnetization, and $\mathbf{H}_{eff}$ is the effective magnetic field acting on $\mathbf{M}$. Since at a fixed temperature much lower than the Curie temperature the saturation magnetization may be regarded as a constant quantity, the LLG equation can be reduced to $d\mathbf{m}/dt = -\gamma^* \mathbf{m} \times \mathbf{H}_{eff} - \alpha \gamma^* \mathbf{m} \times (\mathbf{m} \times \mathbf{H}_{eff})$, where $\mathbf{m} = \mathbf{M}/M_s$ and $\gamma^* = 1/(1+\alpha^2)$.

The effective field $\mathbf{H}_{eff}$ involved in the LLG equation is the sum of the external magnetic field $\mathbf{H}$, field $\mathbf{H}_{dip}$ caused by magnetostatic dipolar interactions between spins, and contributions resulting from the magnetocrystalline anisotropy ($\mathbf{H}_{mca}$), magnetoelastic coupling ($\mathbf{H}_{mel}$), and exchange interaction ($\mathbf{H}_{ex}$). The calculation of $\mathbf{H}_{dip}$ is computationally most time consuming because it requires the summing of magnetic fields created by all spins in the studied ensemble at each spin position. To reduce the simulation time to a reasonable level, we introduce nanoscale computational cells with dimensions much larger than the unit cell size but smaller than the exchange length. The second condition guarantees that the magnetization orientation does not change significantly within an individual cell. Therefore, the introduced cells may be modeled by uniformly magnetized rectangular prisms, and the dipolar field $\mathbf{H}_{dip}$ can be calculated as a sum of magnetic fields created by such prisms. Accordingly, $\mathbf{H}_{dip}$ may be written in the general form as

$$\mathbf{H}_{dip}(\mathbf{r}) = \sum_n \mathbf{N}(\mathbf{r} - \mathbf{r}_n)\mathbf{m}(\mathbf{r}_n), \quad (1)$$

where the summation is carried out over all computational cells, $\mathbf{r}_n$ denote the vectors defining spatial positions of their centers, $\mathbf{m}(\mathbf{r}_n) = \mathbf{m}_n$ is the magnetization direction in the $n$-th cell, and the matrix $\mathbf{N}$ is described by analytical relations [17, 18]. For the numerical computation of the exchange field $\mathbf{H}_{ex}$, it is convenient to use the relation

$$\mathbf{H}_{ex}(\mathbf{r}_n) = \frac{2A}{M_s} \sum_{p=1}^{6} \frac{\mathbf{m}(\mathbf{r}_p) - \mathbf{m}(\mathbf{r}_n)}{d_p^2}, \quad (2)$$

where $A$ is the exchange stiffness coefficient [19], the summation is carried out over the six nearest neighbors of the $n$-th cell, $d_p = |\mathbf{r}_p - \mathbf{r}_n|$, and the differences between the magnetization orientations in neighboring cells are assumed to be smaller than 30° [20]. The remaining contributions to $\mathbf{H}_{eff} = \mathbf{H} + \mathbf{H}_{mca} + \mathbf{H}_{mel} + \mathbf{H}_{ex} + \mathbf{H}_{dip}$ can be found by differentiating the magnetic energy density $F$ written as a polynomial in terms of the magnetization direction cosines $m_i$ in the Cartesian reference frame $(x,y,z)$ [19]. For ferromagnets with a cubic paramagnetic phase, the relation $\mathbf{H}_{eff} = -\partial F/\partial \mathbf{M}$ gives (no summation over repeated indices $i = x,y,z$, $j \neq i$, and $k \neq i,j$)

$$\mathbf{H}_i^{mca} = -\frac{2}{M_s}\left[K_1(m_j^2 + m_k^2) + K_2 m_j^2 m_k^2\right] m_i, \quad (3)$$

$$\mathbf{H}_i^{mel} = -\frac{1}{M_s}\left[2B_1 u_{ij} m_i + B_2(u_{ij} m_j + u_{ik} m_k)\right], \quad (4)$$

where $K_1$ and $K_2$ are the magnetocrystalline anisotropy constants of fourth and sixth order, $B_1$ and $B_2$ are the magnetoelastic coupling constants, and $u_{ij}$ are the lattice strains.

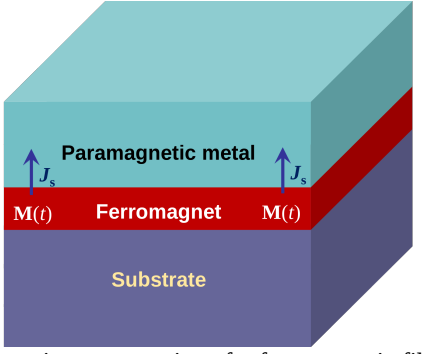

**Fig. 1.** Schematic representation of a ferromagnetic film grown on a nonmagnetic substrate and covered by a thick layer of paramagnetic metal. The magnetization oscillations driven by an elastic wave create a spin current $\mathbf{J}_s$ pumped from the film to the adjacent paramagnetic layer.

Since the magnetization reorientations modify lattice strains, the LLG equation generally should be solved together with the elastodynamic equation of motion [11, 12, 16]. However, the magnetostrictive strains are very small (~$10^{-5}$) so that we can neglect them in comparison with the strains $u_{ij}$ induced by an elastic wave even at $u_{ij} \sim 10^{-3}$. Therefore, the distribution of lattice strains in the film was assumed independent of the magnetic pattern. In our model simulations, two types of standing waves were considered, namely, the longitudinal and transverse waves defined by the relations $u_{xx} = u_{max}\sin(2\pi x/\lambda)\cos(2\pi \nu t)$ and $u_{xz} = u_{zx} = u_{max}\sin(2\pi x/\lambda)\cos(2\pi \nu t)$, respectively. Strictly speaking, strain waves with such simple structure can exist only in a film sandwiched between two elastic half-spaces with the same elastic properties as the film. However, they also represent a reasonable approximation for elastic waves in a relevant material system having the form of a thin ferromagnetic film grown on a nonmagnetic substrate and covered by a thick layer of a paramagnetic metal (see Fig. 1).

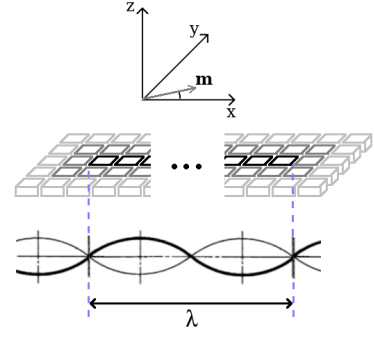

**Fig. 2.** Schematic of a ferromagnetic film divided into nanoscale computational cells for simulations of the magnetization dynamics induced by standing elastic waves. The axes of the rectangular coordinate system ($x$, $y$, $z$) are parallel to the crystallographic axes of the ferromagnet, the unit vector $\mathbf{m}$ shows the initial orientation of the magnetization, and $\lambda$ is the wavelength of the standing wave.

The simulations were performed using a home-made software which operates with a finite ensemble of $N$ computational cells characterized by their spatial positions $\mathbf{r}_n$ and time-dependent unit vectors $\mathbf{m}_n(t)$ ($n = 1,2,…N$). First, the effective fields $\mathbf{H}_{eff}(\mathbf{m}_n)$ acting on vectors $\mathbf{m}_n(t)$ at the moment $t$ are calculated with the aid of Eqs. (1)-(4). Using the computed fields and the known set of $\mathbf{m}_n(t)$, we integrate the LLG equation numerically and determine the magnetization orientations $\mathbf{m}_n$ in all cells at the moment $t + \delta t$. This procedure is repeated until a steady periodic solution for the strain-induced dynamic magnetic pattern is obtained. Since in our case the strain distribution has the form of a standing wave, periodic boundary conditions along the $x$ axis may be introduced for a ferromagnetic film parallel to the $xy$ plane, which enables us to consider only cells situated within one wavelength $\lambda$ (Fig. 2).

The developed computational scheme employs the LLG equation written in the Cartesian coordinates. As the LLG equation is known to be "stiff", the numerical integration is performed using the projective Euler scheme with a fixed integration step $\delta t = 5$ fs, where the condition $|\mathbf{m}| = 1$ is satisfied automatically. To reduce the computation time, the dipolar field $\mathbf{H}_{dip}$ is calculated with the aid of fast Fourier transforms and the convolution theorem. In addition, the magnetic fields of uniformly magnetized prisms are replaced by the fields of point magnetic dipoles at distances exceeding the cell sizes by more than a factor of 50. Based on the symmetry of the problem, the orientation of all vectors $\mathbf{m}_n(t)$ in each chain of cells parallel to the in-plane $y$ axis is taken to be the same in each moment $t$. To test the accuracy of our software, we used it to solve the NIST Standard Problem #4 [21] and found good agreement with the reference.

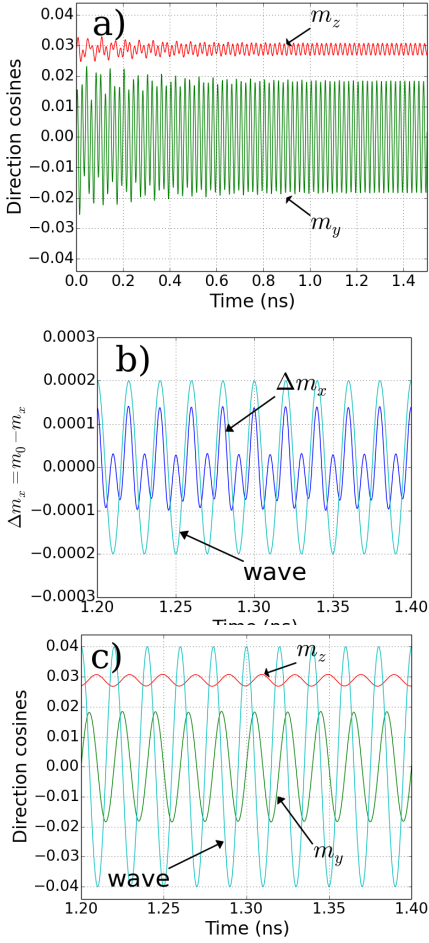

**Fig. 3.** Magnetization dynamics induced by the transverse standing wave with the frequency $\nu \cong 50$ GHz much higher than the resonance frequency $\nu_{res} \cong 9.89$ GHz of unstrained $Fe_{81}Ga_{19}$ film. Panel (a) shows the temporal evolution of the magnetization direction cosines $m_y$ and $m_z$ in the whole simulation including the transient regime. Panels (b) and (c) present the enlarged view of the regime of steady-state magnetization precession. Phase shifts between the periodic variations of the direction cosines $m_y$ and $m_z$ and shear strains in the wave amount to $\pi/2$ and $\pi$, respectively.

The simulations were performed for a 2 nm thick $Fe_{81}Ga_{19}$ film using the following values of the involved material parameters: $M_s = 1321$ emu cm$^{-3}$ [22], $\alpha = 0.017$ [23], $A = 1.8 \times 10^{-6}$ erg cm$^{-1}$ [24], $K_1 = 1.75 \times 10^5$ erg cm$^{-3}$, $K_2 = 0$ [25], $B_1 = -0.9 \times 10^8$ erg cm$^{-3}$, and $B_2 = -0.8 \times 10^8$ erg cm$^{-3}$ [14]. To stabilize the single-domain initial state in the ferromagnetic film, an external magnetic field with the components $H_x = H_z = 500$ Oe was introduced. Since the exchange length $l_{ex} = \sqrt{A/(2\pi M_s^2)}$ [26] of $Fe_{81}Ga_{19}$ is about 4 nm, one computational cell is sufficient in the film thickness direction. Therefore, we employed cells with the dimensions $2 \times 2 \times 2$ nm$^3$ and considered standing elastic waves with wavelengths $\lambda$ equal to even numbers of the 2 nm cell size only. The frequencies of these waves were determined from the dispersion relation $\nu = c_{l,t}/\lambda$, where the phase velocities $c_l = \sqrt{c_{11}/\rho}$ and $c_t = \sqrt{c_{44}/\rho}$ of longitudinal and transverse waves were calculated using the elastic stiffnesses $c_{11} = 1.62 \times 10^{12}$ dyne cm$^{-2}$, $c_{44} = 1.26 \times 10^{12}$ dyne cm$^{-2}$, and density $\rho = 7.8$ g cm$^{-3}$ of $Fe_{81}Ga_{19}$ [27]). For both waves, the strain amplitude $u_{max}$ was set equal to $0.5 \times 10^{-2}$.

### III. MAGNETIZATION DYNAMICS IN ELASTIC WAVES

The magnetization dynamics caused by magnetoelastic coupling should depend on the frequency of elastic wave. Therefore, we carried out simulations for standing waves with various wavelengths, which provide a wide frequency range spanning frequencies below and above the resonance frequency $\nu_{res}$ of coherent magnetization precession in unstrained $Fe_{81}Ga_{19}$ film. This resonance frequency was determined by studying the relaxation of magnetization vector to the equilibrium orientation and found to be about 9.89 GHz at the considered applied magnetic field ($H_x = H_z = 500$ Oe).

All simulations started at the equilibrium magnetic state of unstrained ferromagnetic film, where the uniform magnetization is directed in the $xz$ plane at an angle of $1.65°$ with respect to the film surfaces. This initial state transforms into a nonhomogeneous magnetic pattern just upon the introduction of a strain wave at $t = 0$. After a transition period of the order of 1 ns (~$10^5$ simulation steps), the magnetic dynamics acquires the form of a steady-state magnetization precession. The angular deviations from the initial magnetization direction are maximal at the antinodes of standing waves, where the driving force of magnetoelastic origin has the largest value. Remarkably, the magnetization precession does not vanish at the nodes, where the driving force goes to zero, which is due to cooperative effects caused by magnetostatic and exchange interactions between spins.

Figure 3 shows the temporal evolution of the magnetization orientation at the antinode ($x = \lambda/4$) of the transverse (shear) standing wave with the frequency $\nu \cong 50$ GHz well above the resonance frequency $\nu_{res}$. The most important finding here is that the steady magnetization precession, which sets in after a transient regime comprising about 60 oscillations [Fig. 3(a)], occurs with the frequency of elastic wave [Figs. 3(b) and 3(c)]. This is a nontrivial result because the magnetoelastic components of the effective field $H_x^{\mathrm{mel}} = -(B_2/M_s)u_{xz}m_z$ and $H_z^{\mathrm{mel}} = -(B_2/M_s)u_{zx}m_x$ acting on the magnetization oscillate with higher frequencies owing to periodic variations of the direction cosines $m_z$ and $m_x$. Such oscillations of $\mathbf{H}_{\mathrm{mel}}$ manifest themselves in the appearance of two maxima of $m_x$ during one period of elastic wave [Fig. 3(b)]. This feature results from specific spatial trajectory of the end of magnetization vector, which does not have the form of a planar curve (see Fig. 4).

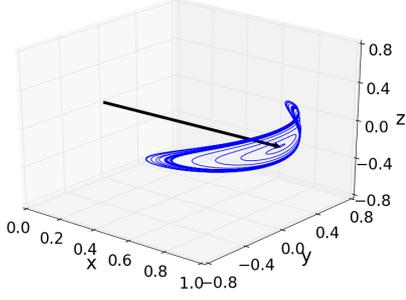

**Fig. 4.** Typical trajectory of the end of magnetization vector at the antinodes of transverse standing waves. The three-dimensional plot presents the full trajectory of the unit vector **m** = **M**/$M_s$ at the wave frequency $\nu \cong 9.88$ GHz. The arrow shows the equilibrium magnetization direction.

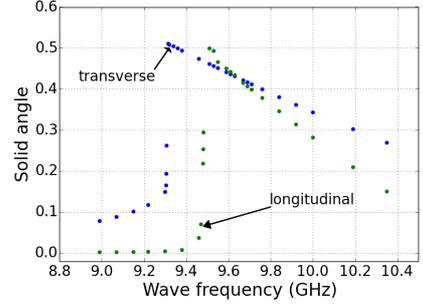

**Fig. 6.** Frequency dependences of the solid angle of steady-state magnetization precession induced in the $Fe_{81}Ga_{19}$ film at the antinodes of transverse and longitudinal standing waves.

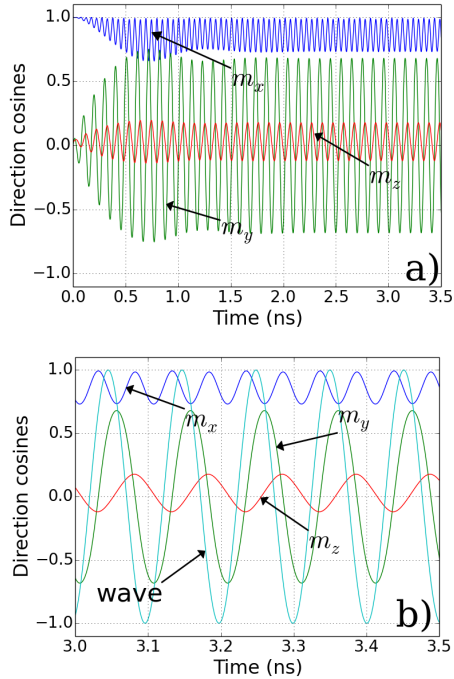

**Fig. 5.** Temporal evolution of the magnetization orientation at the antinode of the transverse standing wave with the frequency $\nu \cong 9.85$ GHz. Panel (a) shows variations of the magnetization direction cosines $m_i$ in the whole simulation including the transient regime, while panel (b) presents the enlarged view of the regime of steady magnetization precession.

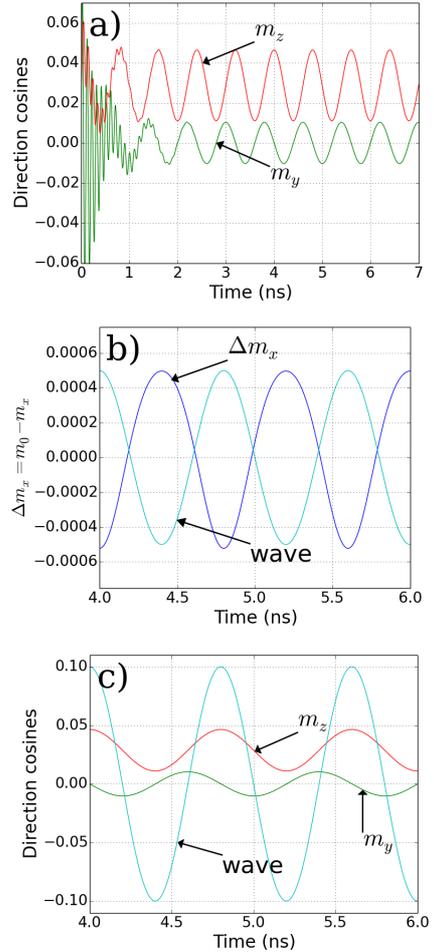

**Fig. 7.** Magnetization dynamics at the antinode of the transverse standing wave with the frequency $\nu \cong 1.25$ GHz. Panel (a) shows the temporal evolution of the magnetization direction cosines $m_y$ and $m_z$ in the whole simulation including the transient regime, while panels (b) and (c) present the enlarged view of the regime of steady magnetization precession ($m_0 = 0.9995$ is the initial value of the direction cosine $m_x$).

When the frequency of elastic wave is reduced down to $\nu_{\mathrm{res}}$, the magnetization oscillations increase drastically (Fig. 5). In this case, the transient regime is distinguished by the presence of intermediate stage, where these oscillations are larger than in the steady-state regime (see Fig. 5(a)). Interestingly, the amplitude of steady magnetization precession reaches maximum at the frequency $\nu_{\max} \cong 9.38$ GHz slightly lower than the resonance frequency of unstrained ferromagnetic film (Fig. 6). At this frequency, the solid angle of magnetization precession exceeds 0.5, but it decreases

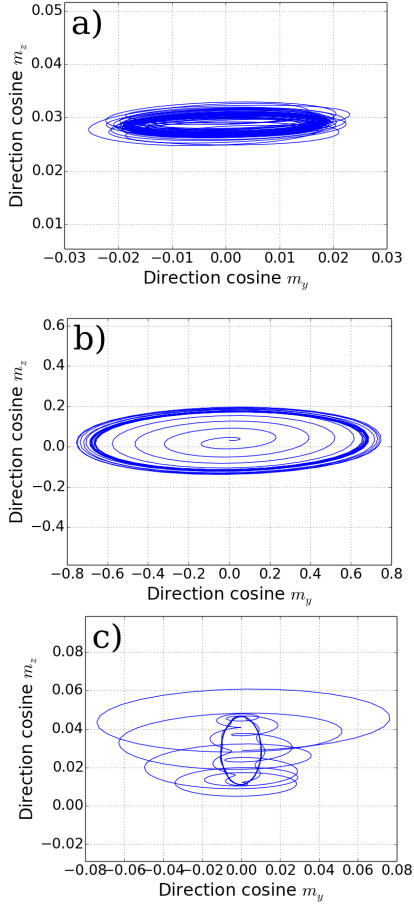

**Fig. 8.** Magnetization trajectories at the antinodes of transverse standing waves projected on the $yz$ plane orthogonal to their wave vectors. Panels (a), (b), and (c) show the projections of the end of the unit vector $\mathbf{m} = \mathbf{M}/M_s$ calculated at the wave frequencies of 50, 9.85, and 1.25 GHz, respectively.

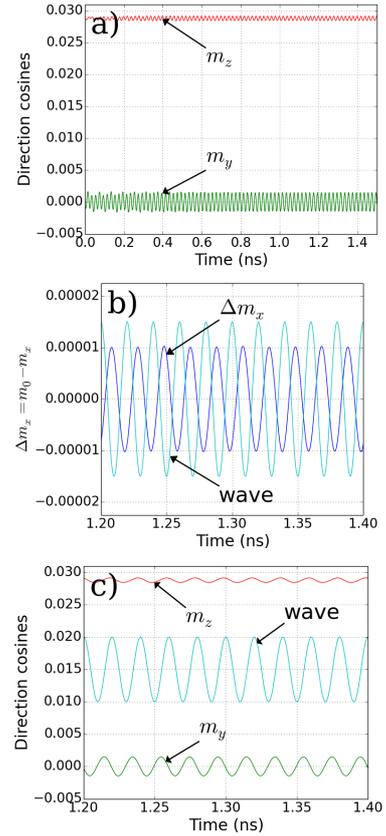

**Fig. 9.** Magnetization dynamics at the antinode of the longitudinal standing wave with the frequency $\nu \cong 50$ GHz excited in the Fe$_{81}$Ga$_{19}$ film. Panel (a) shows the temporal evolution of the magnetization direction cosines $m_y$ and $m_z$ in the whole simulation including the transient regime, while panels (b) and (c) present the enlarged view of the regime of steady-state magnetization precession ($m_0 = 0.9995$ is the initial value of the direction cosine $m_x$).

rapidly at smaller frequencies, falling down to 0.1 already at $\nu \cong 9.15$ GHz.

Figure 7 illustrates the magnetization dynamics induced by the transverse standing wave with the frequency $\nu \cong 1.25$ GHz well below the resonance frequency $\nu_\mathrm{res}$. A novel feature here is the presence of a *double dynamics* in the transient regime [see Fig. 7(a)]. In contrast to the case of $\nu \gg \nu_\mathrm{res}$, the magnetization precesses with the frequency $\nu \cong 10$ GHz $\cong \nu_\mathrm{res}$ much higher than the wave frequency. This fast dynamics is accompanied by slow variations of the precession trajectory following the evolution of elastic wave. In the steady-state regime, the frequency of magnetization precession drops down to the wave frequency [Fig. 7(b)], which is accompanied by a drastic change in the precession trajectory (see Fig. 8).

The magnetization oscillations excited by longitudinal standing waves with three representative frequencies are shown in Figs. 9-11. At the high frequency $\nu \cong 50$ GHz (Fig. 9), the magnetic dynamics is qualitatively similar to the one discussed above for the transverse wave with the same frequency. However, angular deviations from the equilibrium magnetization direction are much smaller in the longitudinal wave despite the fact that the magnetoelastic constants $B_1$ and $B_2$ have almost the same value in the considered case of Fe$_{81}$Ga$_{19}$ alloy [14]. This smaller efficiency of longitudinal waves originates from dissimilar structure of their contribution $H_x^\mathrm{mel} = -2(B_1/M_s)u_{xx}m_x$ to the effective field.

At frequencies close to the resonance frequency $\nu_\mathrm{res}$ of unstrained film, the amplitude of magnetization precession strongly increases (Fig. 10). This is accompanied by significant distortions of the time dependences of direction cosines $m_i$, which remain periodic ($\nu \cong 9.89$ GHz) but cannot be described by simple sine or cosine functions in the steady-state regime. Interestingly, the dependence $m_x(t)$ is distinguished by the presence of two maxima during one wave period, which is similar to the behavior of $m_x$ in the transverse wave with $\nu \gg \nu_\mathrm{res}$ shown in Fig. 3(b). The solid angle of magnetization precession induced by longitudinal waves reaches maximum at the frequency $\nu_\mathrm{max} \cong 9.61$ GHz slightly higher than the most efficient frequency $\nu_\mathrm{max}$ of transverse waves (see Fig. 6).

When the frequency of longitudinal wave is reduced below $\nu_\mathrm{res}$, the magnetic dynamics changes dramatically (Fig. 11). Most importantly, the magnetization precesses with a *variable frequency* strongly exceeding the wave frequency. This feature may be attributed to the dependence of the resonance frequency $\nu_{res}$ of coherent precession on lattice strains [28]. Indeed, the calculation shows that the change of $u_{xx}$ from $-0.005$ to $+0.005$ gives $\nu_{res}$ ranging from 3.44 to 14.45 GHz, which agrees with the frequency range 4.3–15 GHz obtained from micromagnetic simulations. These frequency variations are accompanied by periodic changes of the precession trajectory following the evolution of elastic wave. In contrast to the case of transverse waves, this double dynamics does not disappear in the steady-state regime (see Fig. 11). The magnetization trajectories at three representative frequencies of longitudinal waves are compared in Fig. 12.

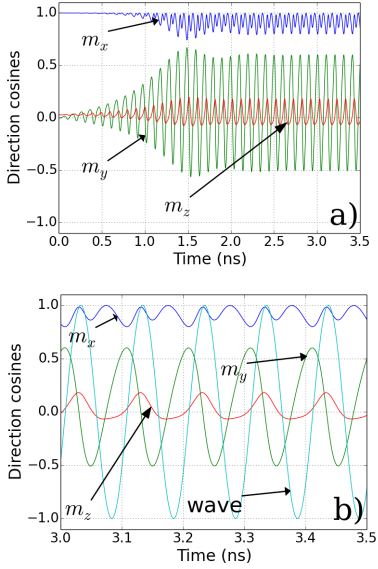

**Fig. 10.** Temporal evolution of the magnetization orientation at the antinode of the longitudinal standing wave with the frequency $\nu \cong 9.89$ GHz. Panel (a) shows variations of the magnetization direction cosines $m_i$ in the whole simulation including the transient regime, while panel (b) presents the enlarged view of the regime of steady magnetization precession.

In conclusion of this section, we consider spatio-temporal distributions of the magnetization oscillations in standing elastic waves. Figures S1 and S2 in the Supplemental Material [29] demonstrate that such elastic excitations generate *standing spin waves* with the same wavelength. Remarkably, the amplitude of magnetization precession does not go to zero at the nodes of these spin waves. Of course, the precession amplitude is always much larger at the antinodes than at the nodes, but the ratio of the amplitude at the antinode to that at the node decreases strongly when the frequency of elastic wave increases from $\nu \ll \nu_{res}$ to $\nu \gg \nu_{res}$. Graphs demonstrated by Figs. S1 and S2 [29] further show that the elastically generated standing spin waves may have very complex structure, especially when exited by a longitudinal elastic wave. In all cases, these spin waves cannot be precisely described by simple relations of the form $m_i(x,t) = m_i^0 + \delta m_i \sin(2\pi x/\lambda)\cos(2\pi\nu t)$. To gain additional information on their structure, we performed the Fourier analysis of the spatial distributions of $m_i$ formed at the moment when the strain at the antinodes of elastic wave reaches its maximum value. The calculations showed that, in the case of the spin wave induced by the transverse elastic wave with the frequency $\nu \ll \nu_{res}$, a term proportional to $\sin(2\pi x/\lambda)$ is sufficient to describe the direction cosines $m_y(x)$ and $m_z(x)$ with a good accuracy. This term also provides the main contribution to $m_y(x)$ and $m_z(x)$ in the spin waves generated by transverse and longitudinal elastic waves with frequencies $\nu \geq \nu_{res}$, but here significant additional contribution (up to 30% of the leading term) is caused by a term proportional to $\sin(6\pi x/\lambda)$. Finally, in the spin wave exited by the longitudinal elastic wave with $\nu \ll \nu_{res}$, the distribution of the direction cosine $m_y(x)$ may be approximated by the sum of terms proportional to $\sin(4\pi x/\lambda)$ and $\sin(20\pi x/\lambda)$, while approximate description of $m_z(x)$ requires terms proportional to $\sin(2\pi x/\lambda)$ and $\sin(8\pi x/\lambda)$.

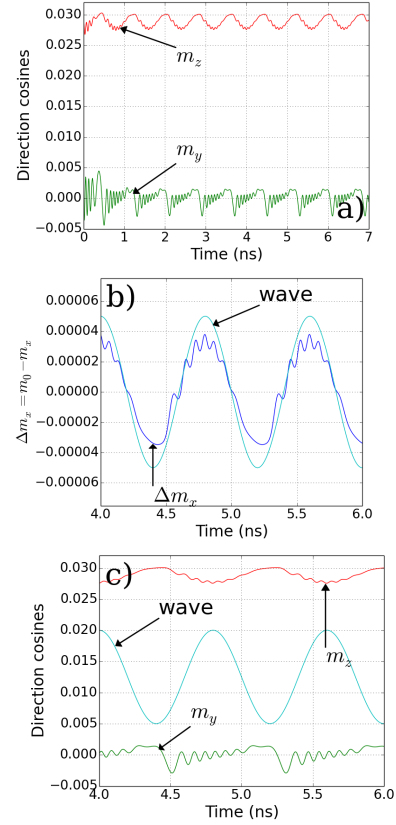

**Fig. 11.** Magnetization dynamics at the antinode of the longitudinal standing wave with the frequency $\nu \cong 1.25$ GHz. Panel (a) shows the temporal evolution of the magnetization direction cosines $m_y$ and $m_z$ in the whole simulation including the transient regime, while panels (b) and (c) present the enlarged view of the regime of steady magnetization precession.

## IV. SPIN PUMPING DRIVEN BY ELASTIC WAVES

When a ferromagnet is in contact with a paramagnetic metal, the magnetization precession in the former leads to a spin pumping into the latter [30]. The spin current density $\mathbf{I}_s$ at the interface can be calculated from the

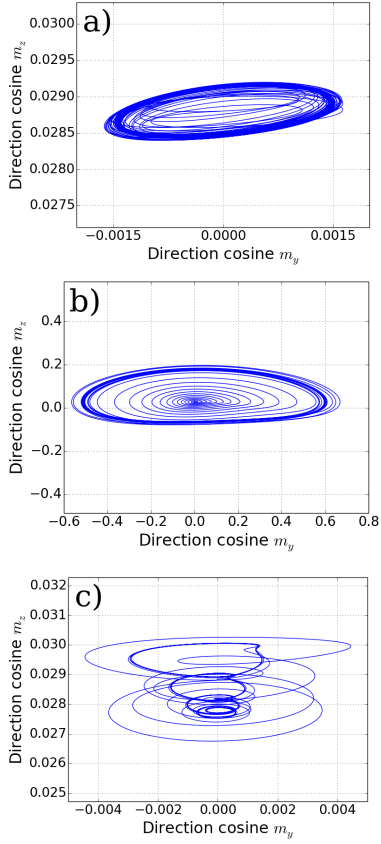

**Fig. 12.** Magnetization trajectories at the antinodes of longitudinal standing waves projected on the *yz* plane orthogonal to their wave vectors. Panels (a), (b), and (c) show the projections of the end of the unit vector **m** = **M**/$M_s$ calculated at the wave frequencies of 50, 9.89, and 1.25 GHz, respectively.

relation [30-32]

$$\mathbf{I}_s = j_s \mathbf{s} = \frac{\hbar}{4\pi}\left(\mathrm{Re}[g^r_{\uparrow\downarrow} - g^t_{\uparrow\downarrow}]\mathbf{m}\times\frac{d\mathbf{m}}{dt} + \mathrm{Im}[g^r_{\uparrow\downarrow} - g^t_{\uparrow\downarrow}]\frac{d\mathbf{m}}{dt}\right), \quad (5)$$

where **s** is the unit vector of the spin-current polarization, and $g^r_{\uparrow\downarrow}$, $g^t_{\uparrow\downarrow}$ are the complex reflection and transmission spin mixing conductances per unit contact area [33, 34]. According to Eq. (5), the magnetization precession induced by elastic waves should create a spin current comprising dc and ac components. The results of our micromagnetic simulations enable us to calculate both dc and ac spin currents generated by standing elastic waves. Since the first-principles studies of spin mixing conductances [34] show that $g^t_{\uparrow\downarrow}$ and the imaginary part of $g^r_{\uparrow\downarrow}$ should be negligible for the 2 nm thick ferromagnetic film considered in this work, in our calculations we used the approximate relation

$$\mathbf{I}_s \simeq \frac{\hbar}{4\pi}\left(\mathrm{Re}[g^r_{\uparrow\downarrow}]\mathbf{m}\times\frac{d\mathbf{m}}{dt}\right). \quad (6)$$

Figure 13(a) shows the time dependence of the spin current created at the antinode of the transverse standing wave with the frequency close to $\nu_{\mathrm{res}}$. It can be seen that, in the steady-state regime, all three components $s_i$ of

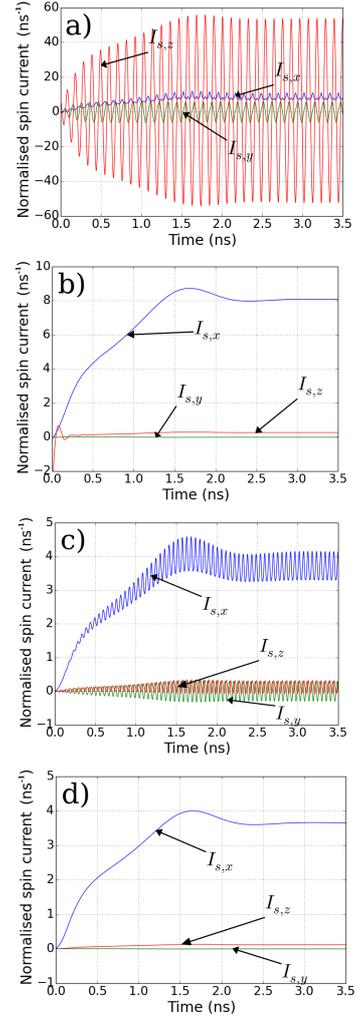

**Fig. 13.** Time dependence of the spin current generated by the $Fe_{81}Ga_{19}$ film subjected to the transverse elastic wave with the frequency of 9.38 GHz. The components $I^s_i$ of the spin-current density are normalized by the quantity $(\hbar/4\pi)Re[g^r_{\uparrow\downarrow}]$. Panels (a) and (b) show $I^s_i$ at the antinode of the standing wave, while panels (c) and (d) present the current density averaged over one wavelength of this elastic wave. The direct results of calculations are given in panels (a) and (c), whereas panels (b) and (d) show mean current densities obtained after filtering out high-frequency oscillations.

the spin-current polarization **s** oscillate with the wave frequency. Interestingly, the oscillations of the spin-current component $I^s_z(t)$ have much larger amplitude than the oscillations of $I^s_y(t)$ and $I^s_x(t)$. However, only the projection of the spin current on the *x* axis has significant mean value. To determine mean values $\overline{I^s_i}(t)$ of all three spin-current components, we averaged $I^s_i(t)$ during the time period comprising several oscillations [see Fig. 13(b)]. It was found that in the steady-state regime $\overline{I^s_x} \approx (2\hbar/\pi)\,\mathrm{Re}[g^r_{\uparrow\downarrow}]\,\mathrm{ns}^{-1}$, $\overline{I^s_z} \approx (0.05\hbar/\pi)\,\mathrm{Re}[g^r_{\uparrow\downarrow}]\,\mathrm{ns}^{-1}$ while $\overline{I^s_y}$ is negligible. These results can be explained by the fact that, according to Eq. (6), the dc component of the spin current should be parallel to the axis of magnetization precession. In the considered case, the latter is close to the equilibrium magnetization in the unstrained film, which is almost parallel to the *x* axis,

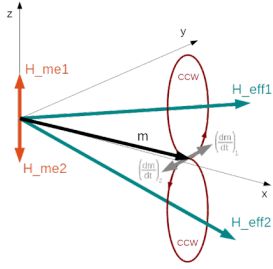

**Fig. 14.** Schematic representation of magnetization precessions in two halves of the standing elastic wave. Despite the fact that the magnetoelastic contribution $\mathbf{H}_{mel}$ of the effective field $\mathbf{H}_{eff}$ has opposite signs in the first and second halves of the standing wave, the magnetization precesses in the same counterclockwise direction in both regions.

having additional small projection on the *z* axis only.

Since the quantity relevant to experimental measurements is the spin current produced by a macroscopic section of the film, we calculated the average current density $\langle I_i^s \rangle$ pumped from the film region corresponding to one wavelength $\lambda$. Figure 13(c) demonstrates that this averaging strongly reduces oscillations of the *y* and *z* components of the spin current. The filtering of high-frequency oscillations (see Fig. 13(d)) further shows that the *x* component retains significant mean value $\langle \overline{I_z^s} \rangle \approx (0.8\hbar/\pi)\,\mathrm{Re}[g_{\uparrow\downarrow}^r]\,\mathrm{ns}^{-1}$, which is only about two times smaller than $\overline{I_x^s}$ at the antinode of transverse standing wave. This feature is due to the fact that the magnetization precesses in the counter-clock-wise direction everywhere despite opposite signs of the driving field $\mathbf{H}_{mel}$ in two halves of the standing wave (see Fig. 14).

The results obtained for the spin currents driven by the longitudinal elastic wave with the frequency close to $\nu_{res}$ are shown in Fig. 15. It can be seen that they are essentially similar to the results discussed above, but there are two interesting distinctions. First, the mean values $\overline{I_y^s}$ and $\langle \overline{I_y^s} \rangle$ of the *y* component are not negligible in the steady-state regime, being close to those of the *z* component of the spin current (see Fig. 15(b) and (d)). This feature is caused by the fact that the axis of magnetization precession driven by the longitudinal wave has a nonzero projection on the *y* axis. Second, in contrast to the case of transverse wave, the averaging of the pumped spin current over the wavelength $\lambda$ leaves the amplitude of the $\langle \overline{I_z^s} \rangle$ oscillations larger than the $\langle \overline{I_x^s} \rangle$ component (compare panels (c) of Figs. 13 and 15). Nevertheless, the mean value $\langle \overline{I_x^s} \rangle \approx (0.6\hbar/\pi)\,\mathrm{Re}[g_{\uparrow\downarrow}^r]\,\mathrm{ns}^{-1}$ of the averaged *x* component remains much larger than the mean value $\langle \overline{I_z^s} \rangle \approx (0.025\hbar/\pi)\,\mathrm{Re}[g_{\uparrow\downarrow}^r]\,\mathrm{ns}^{-1}$ of the *z* one.

Using the theoretical result $(e^2/h)\,\mathrm{Re}[g_{\uparrow\downarrow}^r] \approx 4.66 \times 10^{14}\,\Omega^{-1}\mathrm{m}^{-2}$ obtained for the reflection spin mixing conductance of the Fe/Au interface by first-principles calculations [34], we estimated numerical values of the spin currents pumped from the dynamically strained $Fe_{81}Ga_{19}$ film into adjacent Au layer. In particular, the calculations give

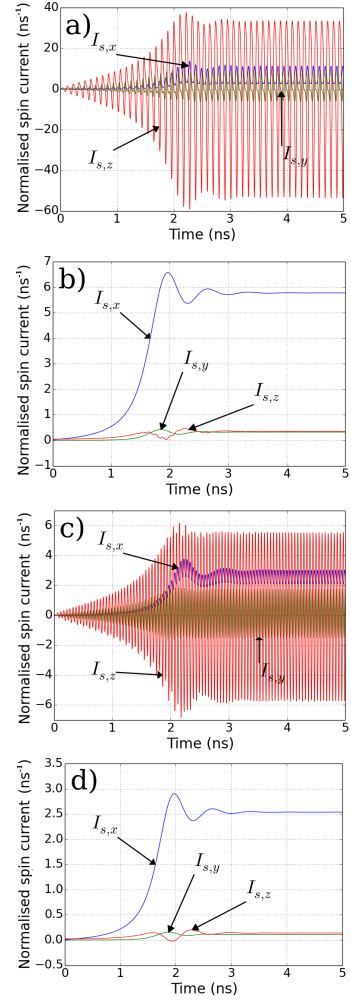

**Fig. 15.** Time dependence of the spin current generated by the $Fe_{81}Ga_{19}$ film subjected to the longitudinal elastic wave with the frequency of 9.61 GHz. The components $I_i^s$ of the spin-current density are normalized by the quantity $(\hbar/4\pi)\,\mathrm{Re}[g_{\uparrow\downarrow}^r]$. Panels (a) and (b) show $I_i^s$ at the antinode of the standing wave, while panels (c) and (d) present the current density averaged over one wavelength of this elastic wave. The direct results of calculations are given in panels (a) and (c), whereas panels (b) and (d) show mean current densities obtained after filtering out high-frequency oscillations.

$\langle \overline{I_x^s} \rangle/\hbar \approx 3.4 \times 10^{27}\,\mathrm{s}^{-1}\mathrm{m}^{-2}$ for the mean current density generated by the transverse elastic wave with the frequency $\nu$ = 9.38 GHz and $\langle \overline{I_x^s} \rangle/\hbar \approx 2.4 \times 10^{27}\,\mathrm{s}^{-1}\mathrm{m}^{-2}$ for the longitudinal wave with $\nu$ = 9.61 GHz. These estimates render possible to evaluate the dc charge current created in the Au layer by the pumped spin current due to the inverse spin Hall effect. The density $\mathbf{I}_c$ of charge current is given by the relation $\mathbf{I}_c = \alpha_{SH}(2e/\hbar)(\mathbf{e}_s \times \mathbf{I}_s)$, where $\alpha_{SH}$ is the spin Hall angle, *e* denotes the elementary positive charge, and $\mathbf{e}_s$ is the unit vector in the spin current direction [32]. Hence, the charge current is orthogonal to the spin current and almost parallel to the *y* axis in our case. Taking $\alpha_{SH} \approx$ 0.0035 for Au [35], we find the current density $\langle \overline{I_y^c} \rangle_{z=0}$ generated at the interface by the transverse and longitudinal waves to be about $3.9\times10^6$ and $2.7\times10^6$ A m$^{-2}$, respectively. Since the injected spin current decays in the normal metal due to spin relaxation and diffusion, the density of charge current falls down with the distance from the interface [36]. The total charge current in the normal metal layer of width $w_N$ and thickness $t_N$ can be found from the relation

$$\langle \overline{J_y^c} \rangle = \langle \overline{I_y^c} \rangle_{z=0}\, \xi_{sd} w_N \frac{\cosh[t_N/\xi_{sd}] - 1}{\sinh[t_N/\xi_{sd}]}, \qquad (7)$$

where $\xi_{sd}$ is the spin diffusion length. Taking $\xi_{sd}$ = 35 nm for Au [36] and assuming $w_N$ = 10 μm, we obtain the total charge current generated in the Au layer with $t_N > 5\, \xi_{sd}$ by the considered standing elastic waves to be about 1 μA, which can be readily measured experimentally.

## V. CONCLUDING REMARKS

In this work, we carried out micromagnetic simulations of the inhomogeneous magnetization dynamics induced in a ferromagnetic material by elastic waves. In contrast to the preceding analytical treatments of the problem [11, 13], our calculations do not involve the assumption of small deviations from the equilibrium magnetization direction. Our approach is based on the numerical solution of the LLG equation comprising the damping term and the effective magnetic field with all relevant contributions, such as those resulting from the magnetoelastic coupling between magnetization and lattice strains, exchange interaction, and magnetocrystalline anisotropy. Furthermore, to describe correctly the practically important case a thin ferromagnetic film, we accurately calculated the magnetostatic dipolar interactions in a ferromagnetic slab of finite thickness. This enabled us to simulate the magnetic dynamics induced by transverse and longitudinal standing waves generated in the 2 nm thick Fe$_{81}$Ga$_{19}$ film sandwiched between two elastic half-spaces. Both the transient and steady-state regimes of magnetization oscillations are described in detail.

The simulations showed that elastic waves induce strongly inhomogeneous magnetization precession, which acquires maximum amplitude at the antinodes of standing waves (near the antinodes in the case of transverse wave with $\nu \gg \nu_{res}$). This amplitude increases drastically near the resonance frequency $\nu_{res}$ of the unstrained ferromagnetic film (Fig. 6), which agrees with the general theoretical predictions [11, 12]. In the steady-state regime, the frequency of magnetization oscillations equals that of the elastic wave, except for the case of longitudinal waves with frequencies well below $\nu_{res}$, where the magnetization precesses with a variable frequency strongly exceeding the wave frequency (Fig. 11(c)).

The spatio-temporal distributions of magnetization oscillations in the considered elastic waves have the form of standing spin waves with the same wavelength $\lambda$ (see Supplemental Material [29]). Remarkably, the amplitude of magnetization precession does not go to zero at the nodes of these spin waves, although angular deviations from the equilibrium magnetization direction here are much smaller than at the antinodes. As a result, the elastically generated standing spin waves cannot be precisely described by simple relations of the form $m_i(x,t) = m_i^0 + \delta m_i \sin(2\pi x/\lambda)\cos(2\pi \nu t)$ even when the magnetization oscillates with the frequency $\nu$ of elastic wave. The simplest structure has the spin wave generated by the transverse elastic wave with $\nu \ll \nu_{res}$, whereas the longitudinal wave with the frequency well below $\nu_{res}$ creates spin wave with the most complex structure.

Using the results obtained for the magnetic dynamics induced by elastic waves, we also calculated the spin currents that can be pumped from the dynamically strained Fe$_{81}$Ga$_{19}$ film into adjacent layer of paramagnetic metal. It was found that both transverse and longitudinal standing waves with the frequency close to $\nu_{res}$ create spin currents comprising significant dc and ac components. Interestingly, the spin polarization of the dc component is not exactly parallel to the equilibrium magnetization direction in the steady-state regime. The calculations of the transverse charge current, which is created by the spin current via the inverse spin Hall effect, showed that the charge current has high density at the interface and can be easily measured experimentally. Our theoretical predictions may be useful for the development of spin injectors driven by elastic waves.

## ACKNOWLEDGMENT

This work was supported by the Government of the Russian Federation through the program P220 (Project No. 14.B25.31.0025; leading scientist A. K. Tagantsev).

# References

[1] M. Farle, *Rep. Prog. Phys.* **61**, 755 (1998).
[2] J. C. Slonczewski and J. Z. Sun, *J. Magn. Magn. Mater.* **310**, 169 (2007).
[3] A. Brataas, A. D. Kent, and H. Ohno, *Nature Materials* **11**, 372 (2012).
[4] W. H. Rippard, M. R. Pufall, S. Kaka, S. E. Russek, and T. J. Silva, *Phys. Rev. Lett.* **92**, 027201 (2004).
[5] A. V. Scherbakov, A. S. Salasyuk, A. V. Akimov, X. Liu, M. Bombeck, C. Brüggemann, D. R. Yakovlev, V. F. Sapega, J. K. Furdyna, and M. Bayer, *Phys. Rev. Lett.* **105**, 117204 (2010).
[6] K. Uchida, H. Adachi, T. An, T. Ota, M. Toda, B. Hillebrands, S. Maekawa, and E. Saitoh, *Nature Materials* **10**, 737 (2011).
[7] M. Weiler, L. Dreher, C. Heeg, H. Huebl, R. Gross, M. S. Brandt, and S. T. B. Goennenwein, *Phys. Rev. Lett.* **106**, 117601 (2011).
[8] M. Weiler, H. Huebl, F. S. Goerg, F. D. Czeschka, R. Gross, and S. T. B. Goennenwein, *Phys. Rev. Lett.* **108**, 176601 (2012).
[9] N. Akulov, *Z. Phys.* **52**, 389 (1928).
[10] T. L. Linnik, A. V. Scherbakov, D. R. Yakovlev, X. Liu, J. K. Furdyna, and M. Bayer, *Phys. Rev. B* **84**, 214432 (2011).
[11] A. Akhiezer, V. Bar'iakhtar, and S. Peletminski, J. *Exptl. Theoret. Phys. (U.S.S.R.)* **35**, 228 (1958).
[12] C. Kittel, *Phys. Rev.* **110**, 836 (1958).
[13] A. Kamra, H. Keshtgar, P. Yan, and G. E. W. Bauer, *Phys. Rev. B* **91**, 104409 (2015).
[14] J. B. Restorff, M. Wun-Fogle, K. B. Hathaway, A. E. Clark, T. A. Lograsso, and G. Petculescu, *J. Appl. Phys*. **111**, 023905 (2012).
[15] D. Afanasiev, I. Razdolski, K. M. Skibinsky, D. Bolotin, S. V. Yagupov, M. B. Strugatsky, A. Kirilyuk, Th. Rasing, and A. V. Kimel, *Phys, Rev. Lett*. **112**, 147403 (2014).
[16] Y. V. Gulyaev, I. E Dikshtein, and V. G Shavrov, *Phys.-Usp.* **40**, 701 (1997).
[17] A. J. Newell, W. Williams, and D. J. Dunlop, *J. Geophys. Res.: Solid Earth* **98**, 9551 (1993).
[18] M. J. Donahue, *Accurate computation of the demagnetization tensor*, http://math.nist.gov/~MDonahue/talks/hmm2007-MBO-03-accurate_demag.pdf
[19] C. Kittel, *Rev. Mod. Phys*. **21**, 541 (1949).
[20] A. Vansteenkiste, J. Leliaert, M. Dvornik, M. Helsen, F. Garcia-Sanchez, and B. Van Waeyenberge, *AIP Advances* **4**, 107133 (2014).
[21] See http://www.ctcms.nist.gov/~rdm/std4/spec4.html
[22] J. Atulasimha and A. B Flatau, *Smart Mater. Struct*. **20**, 043001(2011).
[23] J. V. Jager, A. V. Scherbakov, T. L. Linnik, D. R. Yakovlev, M. Wang, P. Wadley, V. Holy, S. A. Cavill, A. V. Akimov, A. W. Rushforth, and M. Bayer, *Appl. Phys. Lett.* **103**, 032409 (2013).
[24] K. S. Narayan, *Modelling of Galfenol nanowires for Sensor Applications*, Master Thesis (University of Minnesota, 2010), http://conservancy.umn.edu/bitstream/handle/11299/93191/Krishnan_Shankar_May2010.pdf
[25] G. Petculescu, K. B. Hathaway, T. A. Lograsso, M. Wun-Fogle, and A. E. Clark, *J. Appl. Phys.* **97**, 10M315 (2005).
[26] C. A. F. Vaz, J. A. C. Bland, and G. Lauhoff, *Rep. Prog. Phys.* **71**, 056501 (2008).
[27] R. R. Basantkumar, B. J. H. Stadler, R. William, and E. Summers, *IEEE Transactions on Magnetics* **42**, 3102 (2006).
[28] N. A. Pertsev, H. Kohlstedt, and R. Knöchel, *Phys. Rev. B* **84**, 014423 (2011).
[29] See Supplemental Material for graphs of spin waves excited by transverse and longitudinal standing elastic waves.
[30] Y. Tserkovnyak, A. Brataas, G. E.W. Bauer, and B. Halperin, *Rev. Mod. Phys*. **77**, 1375 (2005).
[31] Y. Tserkovnyak, A. Brataas, G. E. W. Bauer, *Phys. Rev. B* **66**, 224403 (2002).
[32] H. J. Jiao and G. E. W. Bauer, *Phys. Rev. Lett*. **110**, 217602 (2013).
[33] A. Brataas, Yu. V. Nazarov, and G. E. W. Bauer, *Phys. Rev. Lett.* **84**, 2481 (2000).
[34] M. Zwierzycki, Y. Tserkovnyak, P. J. Kelly, A. Brataas, and G. E. W. Bauer, *Phys. Rev. B* **71**, 064420 ( 2005).
[35] O. Mosendz, V. Vlaminck, J. E. Pearson, F. Y. Fradin, G. E. W. Bauer, S. D. Bader, and A. Hoffmann, *Phys. Rev. B* **82**, 214403 (2010).
[36] O. Mosendz, J. E. Pearson, F. Y. Fradin, G. E. W. Bauer, S. D. Bader, and A. Hoffmann, *Phys. Rev. Lett.* **104**, 046601 (2010).

# Supplemental material

## Magnetization dynamics and spin pumping induced by standing elastic waves

A. V. Azovtsev and N. A. Pertsev

*Ioffe Institute, 194021 St. Petersburg, Russia*

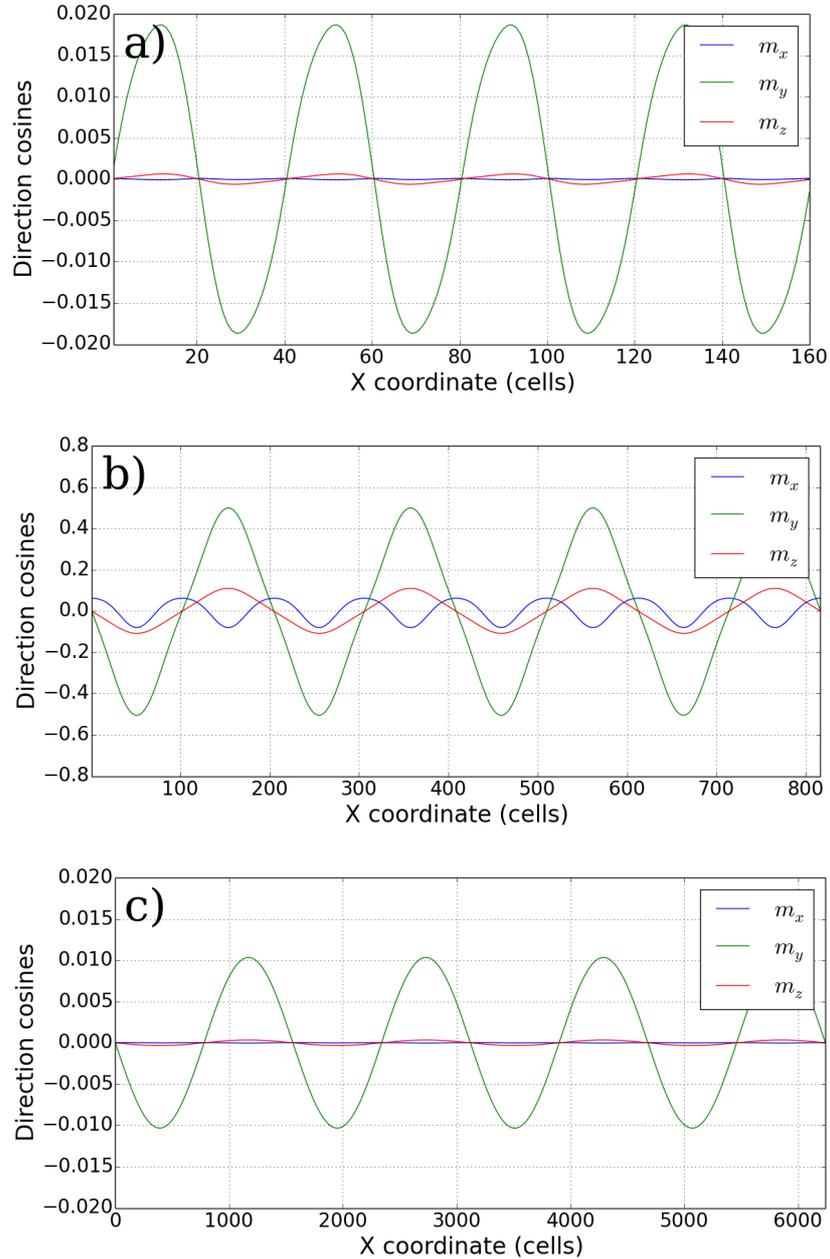

**Fig. 1.** Standing spin waves generated by transverse elastic standing waves in the 2 nm thick $Fe_{81}Ga_{19}$ film (four wavelengths are shown). Panels (a), (b), and (c) demonstrate spin waves excited at the frequency of elastic wave $\nu \cong 50$ GHz, $\nu \cong 9.85$ GHz, and $\nu \cong 1.25$ GHz, respectively.

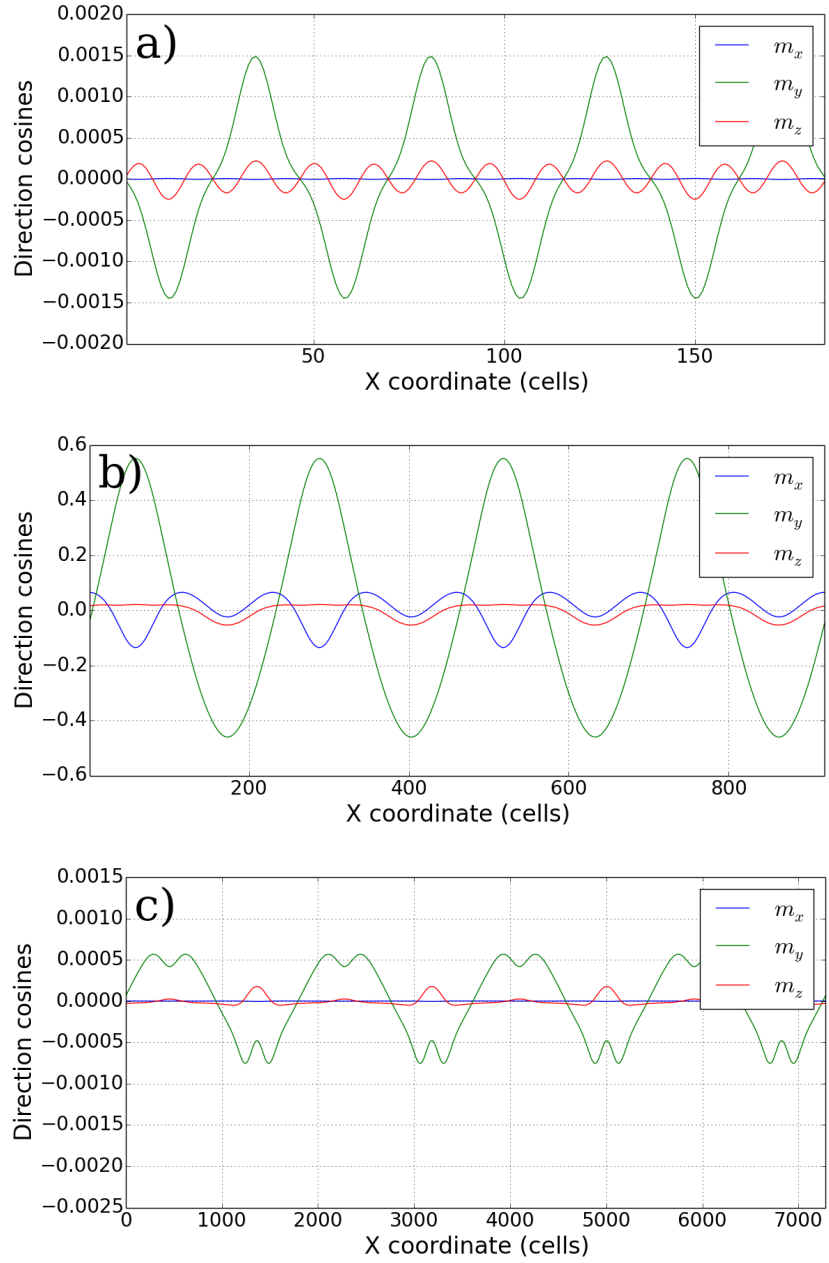

**Fig. 2.** Standing spin waves generated by longitudinal standing elastic waves in the 2 nm thick $Fe_{81}Ga_{19}$ film (four wavelengths are shown). Panels (a), (b), and (c) demonstrate spin waves excited at the frequency of elastic wave $\nu \cong 50$ GHz, $\nu \cong 9.85$ GHz, and $\nu \cong 1.25$ GHz, respectively. The resonance frequency of unstrained $Fe_{81}Ga_{19}$ film is about 9.884 GHz.